\documentclass[11pt]{article}
\usepackage{amsmath}
\usepackage{geometry}                
\usepackage{graphicx}
\usepackage{amssymb}
\usepackage{epstopdf}
\title{\bf Dilaton in a  cold Fermi  gas}
\author{Gordon W. Semenoff\\~~\\{\it \small Department of Physics and Astronomy, University of British Columbia}\\{\it\small
6224 Agricultural Road, Vancouver British Columbia,
Canada V6T 1Z1} }
\date{}                                           

\begin{document}
\maketitle
\begin{abstract}\noindent
 These are the notes for a lecture which I presented at the International Conference on New Frontiers in Physics in Kolymbari, Crete in July, 2018.
They review an idea which posits  a  phase of a two-dimensional system of cold N-component Fermions which
exhibits spontaneously broken approximate scale symmetry when studied in the large N expansion.   Near criticality, the
phase exhibits anomalously small pressure and  large compressibility. 
Some of the consequences of the approximate
scale symmetry, such as the existence of a dilaton and its properties are discussed.\end{abstract}

 \maketitle
\section{Introduction}
\label{intro}
All renormalizable quantum field theories have an approximate scale invariance in that their dimensionful parameters, such
as particle masses, and interaction scales  are small compared to the heaviest scale in the theory, the ultraviolet cutoff.  In this lecture, I wish to discuss the
 question as to whether this approximate scale symmetry
can be spontaneously broken.  Of course, if the scale symmetry is not an exact symmetry, the question is as to whether its spontaneous breaking 
 would play out in a similar way to, say, the breaking of the approximate chiral symmetry of quantum chromodynamics which results
in pseudo-Goldsone bosons, the pions which are light, but not massless.  In the case of scale invariance, the pseudo-Goldstone boson should be a light dilaton. 
The symmetry  breaking would occur when the dynamics of the field theory generates a momentum scale which
  is much larger than the particle masses and other intrinsic scales but still much smaller than the ultraviolet cutoff.  
 
  This idea has been of interest in
 walking technicolor theories for dynamical breaking of the electroweak symmetry in the standard model  \cite{Appelquist:2010gy}-\cite{Hashimoto:2010nw} as
  well as in the idea that the standard model Higgs field could itself be a light dilaton
  \cite{Goldberger:2008zz}-\cite{Csaki:2015hcd}.  Such a scenario is typically difficult to analyze, as dynamical symmetry breaking requires
  strong coupling and it is therefore beyond the reach of perturbation theory. 

    In the following we shall review a non-relativistic field theory model which has a phase with spontaneously broken scale symmetry  \cite{Semenoff:2017ptn}.  
 The model is that of a  cold, non-relativistic, 
 two-dimensional Fermi gas with an attractive delta-function two-body interaction.  Of course, as usual, dynamical symmetry breaking will occur 
  in the strong
 coupling regime.  In order to control the approximations that we shall make, we generalize the Fermion from having two spin components
 to one having $N$ components and a global $U(N)$ symmetry.  Then we shall analyze the theory in the large $N$ limit and the $\frac{1}{N}$ expansion,
 which should be accurate if $N$ is large enough.

We shall study the  quantum critical behaviour 
at the transition where the Fermi gas acquires a density of particles, that is where, as the parameters in the Lagrangian are varied, 
 the particle density crosses from zero to non-zero.

 
  
\section{The Model}
\label{sec-1}

As is usual in the formulation of many-body physics as a quantum field theory, we shall use the grand canonical
ensemble where particles and energy are allowed to flow in and out of the system to a reservoir.  
Then we describe the non-relativistic quantum field theory by using  the 
action, \begin{align}\label{action}
S=\int dtd^2x{\mathcal L}\end{align} with  (imaginary time)  Lagrangian density
 \begin{align}
{\mathcal L}=
\psi_a^*\dot\psi_a+\vec\nabla\psi_a^*\cdot\vec\nabla\psi_a  -\mu \psi_a^*\psi_a
 -\frac{g}{2{\rm N}}(\psi_a^*\psi_a)^2
\label{lagrangian_density}
\end{align}
We shall assume that the  paired indices $a$ run from 1 to N and they are implicitly summed wherever they appear.
The parameter $\mu$ is the chemical potential.   We are working in a system of units where $\hbar=1$ and $2m=1$, where $m$
 is the mass of the Fermions. We have assumed that the interaction between the Fermions  is short-ranged and we have
 approximated it by a two-body contact interaction.   Here, we are using imaginary time since it is most suited to discussing the 
 energies of equilibrium states.  
 
 The Lagrangian has three parameters, the chemical potential $\mu$, the coupling constant, $g$
 and the number of species, $N$.   In order to analyze the model quantitatively, we shall have to take the large $N$ limit \cite{largeN}.  
 Then, in that limit, we will examine the behaviour
 of the system as a function of the remaining parameters, $\mu$ and $g$. The chemical potential breaks scale invariance explicitly. 
   In the infinite $N$ limit, and specifically in two dimensions, $g$ is a tuneable dimensionless parameter.  
   Thus, to get scale invariance, we shall tune $\mu$ to zero and we
   will find that symmetry breaking requires that we also tune the coupling constant $g$ tor a specific value. 
    This symmetry breaking is a direct analog of the Bardeen Moshe Bander phase that is found
   in 
   the large $N$ limit of the O(N) invariant  $\frac{\lambda}{N^2}(\phi^2)^3$ model \cite{Bardeen:1983rv}. That model has been analyzed
   with techniques similar to what we outline here in reference \cite{Omid:2016jve}.  A big difference between that model and the one
   which we shall analyze here is that phase of the  $\frac{\lambda}{N^2}(\phi^2)^3$ model with broken scale symmetry is unstable 
   once $\frac{1}{N}$ corrections to the infinite $N$ limit are taken into  account,  whereas the symmetry breaking phase of the non-relativistic
   Fermi gas that we study here appears to be stable.
   
 When
 the large $N$ limit is relaxed, at the next-to leading order in $\frac{1}{N}$, $g$ becomes a scale-dependent running coupling constant. 
  In the regime where it is attractive, it is asymptotically free and it has a Landau pole in the infrared. It is tuneable in the sense that we can
  set a value for it at the ultraviolet cutoff.   We shall find that the running of the coupling actually helps us in that, 
  rather than a specific fine-tuned value of $g$, we
  find a finite band of renormalization group trajectories for which the spontaneous breaking of the approximate scale invariance occurs. 

The grand canonical thermodynamic potential energy $\Phi(\mu,g,N){\mathcal V} $ and the potential energy density $\Phi(\mu,g,N)$ are defined by the functional integral
\begin{equation}\label{pathintegral0}
 e^{-\Phi(\mu,g,N){\mathcal V}{\mathcal T}}=\int [d\psi_a(x) d\psi^*_a(x)] ~e^{-S[\psi,\psi^*] }
\end{equation}
where ${\mathcal V}$ is the volume of space, ${\mathcal T}$ is the time  and $\psi_a(x)$ are anti-commuting functions. Our discussion will focus exclusively on the
zero temperature limit, where Euclidean time lives on the entire open infinite real line and ${\mathcal T}$ is the length of the real line. 
For the most part, we will be interested in the 
potential density $\Phi(\mu,g,N)$.  

We will study the system described by the functional integral (\ref{pathintegral0}) in the large $N$ expansion.  
The coupling constant in the Lagrangian density (\ref{lagrangian_density}) has been rescaled so that the parameter
$g$ that appears there is held fixed as $N$ is taken as large and the thermodynamic potential is
$$
\Phi(\mu,g,N) = N\Phi_{-1}(\mu,g)+ \Phi_0(\mu,g)+\frac{1}{N}\Phi_1(\mu,g)+\ldots
$$
In the following we will discuss computation of the first two terms in this expansion. 

What we shall do in the following is to observe that, in two dimensions, $g$ is a dimensionless constant and the only parameter in the classical
Lagrangian density with non-zero scaling dimensions is the chemical potential.   If the chemical potential were zero, this quantum field theory would
be scale invariant at the classical level.  We shall then attempt to find a phase of this system where the  density per species of the Fermions, 
\begin{equation}\label{rho}
\rho = \frac{1}{  N}\sum_{a=1}^{  N}<\psi_a^*(x)\psi_a(x)>
\end{equation}
remains finite as the chemical potential is tuned to zero.  

To see that this might be possible, let us consider 
the Hamiltonian density 
\begin{align}\label{hamiltonian_density}
{\mathcal H}=\vec\nabla\psi_a^*\cdot\vec\nabla\psi_a  -\mu \psi_a^*\psi_a
 -\frac{g}{2{\rm N}}(\psi_a^*\psi_a)^2
\end{align}
The first term in (\ref{hamiltonian_density}) is the kinetic energy density.  If this were the only term in the Hamiltonian,
 in the open system that we are considering,  all of the
Fermions would be repelled from the system to the reservoir and the density would be zero.   
This is due to a combination of the non-negative kinetic energy of each Fermion and the Pauli exclusion principle which allows no more than
$N$ Fermions to occupy the same kinetic energy state. The resulting  Fermi pressure would have the result that,
  the Fermions would escape to the reservoir and the density would go to zero. 

In order to retain a finite density, one adds the second, chemical potential term in (\ref{hamiltonian_density}).
That term provides a potential
energy incentive for Fermions to remain in the system.  Adding this term creates a state with finite density that is determined by the
chemical potential that is imposed, 
$$
{\rm Quantum~Ideal~gas:} ~~\rho =\left\{\begin{matrix}\frac{\mu}{4\pi}& \mu>0  \cr 0 & \mu\leq0 \cr \end{matrix} \right.
$$
Here, and in all formulae below, we work in two dimensions and $\rho$ is the Fermion density per species defined in equation (\ref{rho}).  
We note that it is this quantity that
is finite at large $N$, and the total density of Fermions is given by $N\rho$. 
In the ideal Fermi gas, where only the fist two terms in (\ref{hamiltonian_density}) are  non-zero, the  density is determined by the above formula.  
From it, we see that, when the chemical potential is non-zero and positive, the density is also non-zero.   
Moreover, the pressure of this quantum  ideal gas is equal to 
$$
{\rm Quantum~Ideal~gas:} ~~P =N \frac{  \mu^2}{8\pi }=N \frac{\rho^2}{2}4\pi
$$
and the compressibility is
$$
{\rm Quantum~Ideal~gas:} ~~\kappa =\frac{1}{N\rho^2} \frac{d\rho}{d\mu}~,~\kappa_0 = \frac{4\pi}{ N\mu^2} = \frac{N}{4\pi( N\rho)^2}
$$ 
If we ask, in this trivial case, about the quantum critical behaviour where the density crosses from zero to non-zero, this occurs
at the point $\mu=0$ and, at that point, $P=0$ and $\kappa_0$ diverges.  In a trivial sense, the system is scale invariant at this point. 

Now, let us  turn on  the last term in the Hamiltonian density (\ref{hamiltonian_density}).  It is an attractive interaction and its effect should
also be to increase the density.   Our central question here is whether
this interaction, when it is made sufficiently strong, can do the job of the chemical potential  and whether it can eventually replace the chemical potential 
as the agent which retains  Fermions in the system so they have non-zero density, even in the absence of chemical potential.   The answer to this
question will be yes.

\section{Infinite $N$}

At the infinite N limit, we shall find   that the grand canonical potential is equal to  
\begin{eqnarray}\label{infinite_N}
\Phi(\mu,g,N) =~ {\rm Inf}_\rho~N \left[\frac{g}{2}\rho^2 - \frac{1}{8\pi}(\mu+g\rho)^2\theta(\mu+g\rho) +{\mathcal O}(1/N)\right] 
\end{eqnarray}
where ${\rm Inf}_\rho$ indicates that we must find the infimum as a function of $\rho$ of the expression to the right of it. 
The first term in equation (\ref{infinite_N}) comes from the attractive interaction and the second term is from the degeneracy pressure.
The equation for the minimum of the function on the right-hand-side of equation (\ref{infinite_N}),  
determines the physical value of the density for a given value of the chemical potential, 
\begin{align}\label{lng-1}
{\rm Large~N~gas:}~~\rho = 
\left\{   \begin{matrix} 
\frac{\mu} {4\pi-g}   +{\mathcal O}(1/N)  &  0\leq g<4\pi ~,~\mu\geq 0    \cr  
{\rm undetermined} & g=4\pi~,~\mu=0 \cr
0 &  \mu\leq 0~,~ 0\leq g<4\pi    \cr 
\infty & g>4\pi \cr
\end{matrix}\right.
\end{align}
The accuracy of this formula is controlled by the small parameter $\frac{1}{N}$ and it becomes exact in the infinite $N$ limit.  This allows us to take 
the coupling $g$ to large values. 
  We see
that we can indeed tune $\mu$ to zero and keep $\rho$ non-zero in (\ref{lng-1}), if we simultaneously tune $g$ to $4\pi$, a strong-coupling fixed point.  We shall
denote the value of $g$ at the fixed point  by $g^*=4\pi+{\mathcal O}(1/N)$ where we anticipate that it will be corrected at large $N$  (and we will eventually 
find the corrections to order $\frac{1}{N}$).  To look at the quantum phase transition it is better to go to the canonical ensemble where the density is fixed and
the chemical potential is the derived quantity, and 
\begin{align}\label{lng-11}
{\rm Large~N~gas:}~~\mu = 
\left\{   \begin{matrix} 
 ( {4\pi-g})\rho      +{\mathcal O}(1/N)    & \rho\geq  0~,~0\leq g<4\pi   \cr  
0 &   g=4\pi~,~{\rm any~}\rho\geq0  \cr
\end{matrix}\right.
\end{align}
The quantum phase transition lies on either of two quantum critical lines in the $\rho$ versus $g$ plane,  \begin{align}
\begin{matrix}
{\rm I:}  &  \rho=0 & 0\leq g<4\pi \cr 
{\rm II:} & \rho > 0 & g=4\pi   \cr
\end{matrix}
\label{scale_invariant_region}
\end{align}
On both of lines I and II, the system is scale invariant. On segment I, this is trivial as the chemical potential and the density are both zero there.  
Segment II is more interesting as, even though the chemical potential vanishes, and the underlying theory is scale invariant, the density is nonzero
and the scale symmetry is spontaneously broken.  
It has the
further interesting properties that the pressure, given by 
\begin{align}\label{lng-2}
{\rm Large~N~gas:} ~~P =N \frac{ \rho^2}{2}(4\pi-g)   +{\mathcal O}(N^0) 
\end{align}
vanishes on the entire critical line and the compressibility
\begin{align}\label{lng-3}
{\rm Large~N~gas:} ~~\kappa = \kappa_0 \left[ \frac{4\pi}{4\pi-g}   +{\mathcal O}(1/N) \right]
\end{align}
diverges there.  

It is curious that scale invariance implies vanishing pressure.   We can see this from a different point of view. 
Translation invariance of the model (\ref{lagrangian_density}) implies the existence of a
conserved stress-energy tensor  $\Theta_{\mu\nu}(x)$ obeying the continuity equations
$$
\frac{\partial}{\partial t}\Theta_{tt}(\vec x,t) + \frac{\partial}{\partial x^a}\Theta_{at}(\vec x,t)=0
~,~
\frac{\partial}{\partial t}\Theta_{tb}(\vec x,t) + \frac{\partial}{\partial x^a}\Theta_{ab}(\vec x,t)=0
$$
(where, here, $a,b,...$ denote the spatial indices). 
Then, when the chemical potential is set to zero, 
conservation of the Noether current for scale symmetry,
$$
 {\mathcal J}_\mu = \left(2t\Theta_{tt}+x_b\Theta_{tb}, 2t\Theta_{at}+x_b\Theta_{ab}\right)
 $$
 implies an identity for the trace of the stress-energy tensor, 
$$2\Theta_{tt}(x)+\sum_a\Theta_{aa}(x)=0$$ 
The expectation value of this equation  in {\it any} translation invariant and isotropic state of the system 
gives a relationship between the pressure and the internal energy
density of the system in that state \cite{Nishida:2007pj} \cite{Golkar:2014mwa}, \begin{align}\label{P=U}P=U\end{align}  

The pressure $P$ and the internal energy density $U$ are also related to the thermodynamic potential $\Phi(\mu,g,N)$
that is computed by (\ref{pathintegral0}) where we note that $\Phi(\mu,g,N)$ is a function of intensive variables only,
so that the thermodynamic identities 
$$
\left. P=-\frac{\partial}{\partial V}[\Phi(\mu,g,N) V]\right|_{\mu,g}=-\Phi(\mu,g,N)
$$
and 
$$
\Phi=U-\mu(N\rho)
$$
 When we set $\mu=0$  this equation yields 
 
 \begin{align}\label{P=-U}
 P=-U
 \end{align}
  and, therefore, that we can have scale invariance only if
 both equations (\ref{P=U}) and (\ref{P=-U}) are satisfied, that is if both
 $$P=0~~{\rm and}~~U=0$$  These equations are obeyed by the state with zero density. 
 If the density is not zero,  the latter condition, $U=0$
 implies that
 the potential energy due to interactions must compensate the kinetic energy of the Fermi gas.  
Clearly this can only happen with a strong attractive interaction.  We have seen that  is so, $g=4\pi$ in the nontrivial scale invariant phase.   
We see from equation (\ref{lng-2}) that, when we set $g$ to $4\pi$, the pressure $P$ vanishes for any value 
of the density $\rho$. 
Moreover, the vanishing pressure, for any value of the density implies infinite compressibility, which we also see in equation (\ref{lng-3})
when we set $g$ to $4\pi$. .

 Before we go on to discuss the next-to-leading order, let us pause to consider the collective behaviour
 at the leading order at large N.   
 For that, we analyze the propagator of the auxillary $\sigma$-field which we shall introduce in the following sections.  
 As we shall discuss there, a the $\sigma$-field is closely related to the density and its connected two-point function 
 differs from the connected correlation function
 for the density by a constant.  Singularities of the $\sigma$=propagator are therefore also singularities of the $\rho$-propagator. 
 We shall find that the 
 $\sigma$-field propagator is given by  $\Delta$ in equation (\ref{PI}) below.     We therefore
 search for collective modes by studying the singularities of $\Delta(i\omega,k)$
 analytically continued to real frequency $i\omega\to \omega$. 
  The elementary one-loop integral yields the polarization function $\Pi(\omega,k)$ defined in equation (\ref{S}), 
\begin{align}
\Pi(\omega,\vec k)=  \frac{ (\omega+{k}^2+i\epsilon) \sqrt{1-    \frac{ 4(\mu+g\rho){k}^2}  {(\omega+{k}^2+i\epsilon)^2}  } -k^2} {8\pi {k}^2} 
+\frac{ (-\omega+{k}^2-i\epsilon) \sqrt{1-    \frac{ 4(\mu+g\rho){k}^2}  {(-\omega+{k}^2-i\epsilon)^2}  } -k^2} {8\pi {k}^2}  
\label{Linhard}
\end{align}
With suitable care to define the cut singularities of the square roots, this is just  the  Linhard function for the two dimensional   Fermi gas with Fermi energy equal to $\mu+g\rho$.
 When the interaction is attractive and subcritical, that is, when $0\leq g<4\pi $, the equation $$\Delta^{-1}=N[g+g^2\Pi(\omega,\vec k)]=0$$ has no real solution for $\omega$. To see this, we simply note that the above equation  can be re-arranged as
 $$
 \frac{ \omega \left[    \sqrt{1-    \frac{ 4(\mu+g\rho){k}^2}  {(\omega+{k}^2+i\epsilon)^2}  }-
  \sqrt{1-    \frac{ 4(\mu+g\rho){k}^2}  {(-\omega+{k}^2-i\epsilon)^2}  }\right] } {8\pi {k}^2}   
 + \frac{  {k}^2\left[ \sqrt{1-    \frac{ 4(\mu+g\rho){k}^2}  {(\omega+{k}^2+i\epsilon)^2}  } 
+  \sqrt{1-    \frac{ 4(\mu+g\rho){k}^2}  {(-\omega+{k}^2-i\epsilon)^2}  } \right] } {8\pi {k}^2}   =\frac{1}{4\pi}-\frac{1}{g}
$$
where we easily see by inspection, in the regions where the square roots are real, the left-hand-side is positive whereas, when $g<4\pi$, 
the right-hand-side is negative.  If either or both square roots are imaginary, the left-hand-side cannot be real and cannot equal
the right-hand-side.   This is 
consistent with the fact that zero sound is strongly damped in a Fermi liquid with 
an attractive interaction. Zero sound normally comes from the real branch of this function and there is a real solution for the
frequency only when $g$ is negative, that is, for the repulsive Fermi gas.

If we study the above equation in the kinematic region where both of the square roots are imaginary, careful treatment of the branches gives their imaginary parts
opposite signs.  Then, the left-hand-side has a zero precisely when $\omega=0$ (and, we must be  inside the branch cuts  which requires $k^2<4(\mu+g\rho)$ ).
The right-hand-side can also be zero, when the coupling is tuned  to $4\pi$.  We see that, in the critical regime, when $g=4\pi$ and 
$\mu+g\rho>0$, there is a pole in the density-density correlation function of the form $\frac{1}{i\omega}$.  Intresetingly, it does not depend on the
wave-number $k$, except for the fact that it is there only when $k^2$ is smaller than the effective Fermi energy $ {4(\mu+g\rho)}$.
 
Near the critical point, and in the $\omega<<k\sqrt{4\pi\rho}<<4\pi\rho$ regime, 
\begin{align}\label{ss}
\Delta(\omega,k)=\frac{\frac{4\pi}{{\rm N}g^2}\sqrt{\frac{\rho}{\pi}}|\vec k|}{\Gamma(k)-i\omega}
~,~
\Gamma(k)=\left(\frac{4\pi}{g}-1\right)  \sqrt{\frac{\rho}{\pi }}|\vec k| 
\end{align}
This collective mode is the dilaton.  It is  damped when $0\leq g<4\pi$.   When $g>4\pi$ it is a growing rather than damping mode, 
symptomatic of the instability of the theory in that regime. It becomes un-damped exactly at  $g=4\pi$, where the  scale symmetry
of the underlying model becomes exact.  

 At that point, $\Gamma=0$ and the pole is at $\omega=0$. 
 The mode is static and non-propagating.  Remember that the Fourier transform of $\frac{1}{i\omega}$ with retarded boundary conditions
 is just the Heavyside step function $\theta(t)$.   This tells us that, in linear response, the critical system is infinitely plastic in that whatever
 deformation one implements simply stays in place thereafter. 
  
   Symmetry breaking in 
 non-relativistic field theories is somewhat richer than what occurs in their relativistic cousins.   
  In non-relativistic systems, there are different types of Goldstone bosons
 as shown in table~\ref{tab-1}. They are classified as type A and type B and,   which type appear in which system,  has
   been resolved only relatively recently \cite{ngb} \cite{Kapustin:2012cr}
 We might postulate that, since our dilaton can damp, but it cannot propagate, it is a third kind, which we
 could call type C, as we have anticipated in table \ref{tab-1}.
\begin{table}[h]
\centering
\caption{Type A and B Goldstone bosons and the dilaton, a proposed  type C.}
\label{tab-1}       

\begin{tabular}{lll}
\hline \hline
Type & dispersion & example  \\\hline 
A &$\omega\sim v|\vec k|$ & antiferromagnet \\
B & $ \omega\sim \gamma \vec k^2$ & ferromagnet \\
C & $\omega\sim 0$ & dilaton \\ \hline
\end{tabular}
\end{table}
 
We also note that the relative of our dilaton is already known in the study of atoms in two dimensional
 harmonic traps.  There, the trapping potential 
breaks scale symmetry.  However, the algebra which would control conformal symmetry where it there is extended to a dynamical
algebra which includes dilatations and the Hamiltonian.  That algebra can be used to show that the gas of trapped atoms has a non-propagating 
breathing mode with frequency that is twice that of the harmonic trapping potential \cite{Pitaevskii}. 
 If one relaxes the trapping potential, that breathing 
mode becomes our dilaton.  

\section{Approximate scale invariance}

In the classical theory described by the Lagrangian density (\ref{lagrangian_density}), when the space dimension is two, 
and the chemical potential  is set to zero,   the scale transformation of the field 
$$
\psi_a(\vec x, t)\to e^{\sigma}\psi(e^\sigma\vec x, e^{2\sigma} t)
$$
is a symmetry of the action (\ref{action})  when $\mu=0$.   
\begin{figure}[h]
\centering
\includegraphics[width=7cm,clip]{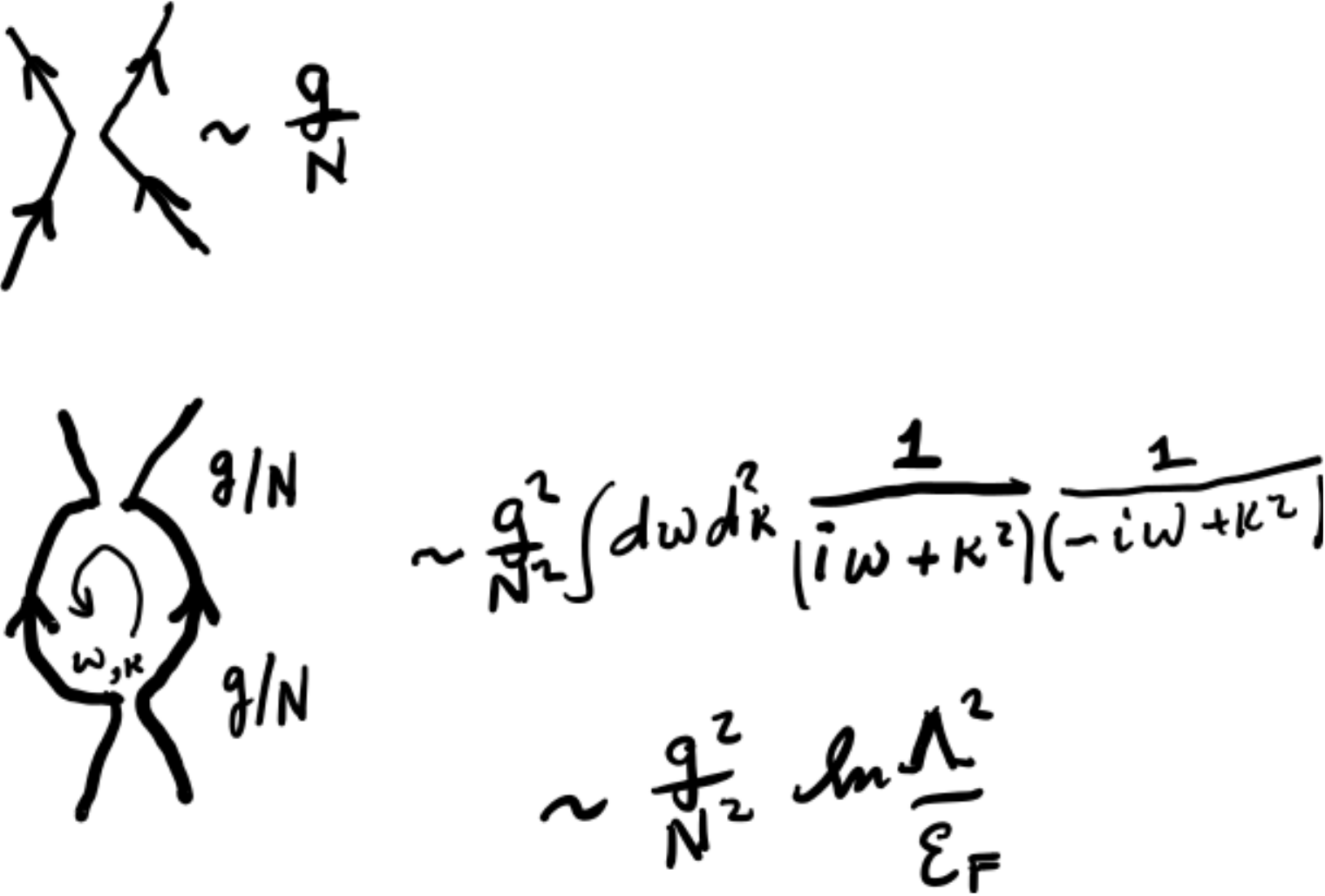}\\
\caption {  The beta function arises from renormalization of the logarithmic divergence that occurs
in the loop correction to Fermion-Fermion scattering, as depicted in the diagram.   Since it involves two vertices
and no summations over Fermion species, it is of order $\frac{g^2}{N^2}$ where  the leading order is $\frac{g}{N}$.
The resulting correction to $g$ and therefore the beta function is thus seen to be of  order $\frac{g^2}{N}$. 
 }   \label{fg-1}  
 \end{figure}  
However, 
it is well-known that quantization of the theory described by (\ref{lagrangian_density}) and (\ref{pathintegral0}) breaks the scale invariance 
with a scale anomaly \cite{anom0}-\cite{an3}. 
The coupling constant $g$ obtains a beta-function \cite{Bergman:1991hf} 
from the renormalization of a logarithmic ultraviolet divergence which occurs
in Fermion-Fermion two-body scattering depicted in figure \ref{fg-1}.

 In the large $N$ expansion, the leading order contribution to the beta function is 
\begin{align}\label{beta}
\beta(g(\Lambda^2)) = \Lambda^2 \frac{d}{d\Lambda^2}g(\Lambda^2)=-\frac{g^2(\Lambda^2)}{8\pi{\rm  N}}+{\mathcal O}(\tfrac{1}{\rm N}^2) 
\end{align}
A nonzero beta function renders $g$  a scale-dependent running coupling, 
 \begin{align}
 g(\mu) = \frac{  g(\Lambda^2) } { 1+\frac{g (\Lambda^2)} {8\pi N} \ln \frac{\mu}{\Lambda^2} }
\label{beta1}
\end{align}
where $\Lambda$ is an ultraviolet cutoff with dimension of 1/distance and $\mu$ is an energy with dimension of 1/distance$^2$.    
The beta function is  small and the scale dependence is weak  in the large N  limit where N$\to\infty$ with $g$ fixed.  
In the strict infinite N limit, the model is scale invariant and the analysis of the phase structure of the model in that limit 
was the considered in the previous section.   There we noted  distinct phenomena, vanishing chemical potential, vanishing pressure and  diverging compressibility, 
which we expect to survive in an approximate sense when the 
$\frac{1}{N}$ corrections are included.  In the following, we shall study this model  at large but finite N.  We will also 
derive and justify some of the formula quoted in the previous section.

\begin{figure}[h]
\centering
\includegraphics[width=7cm,clip]{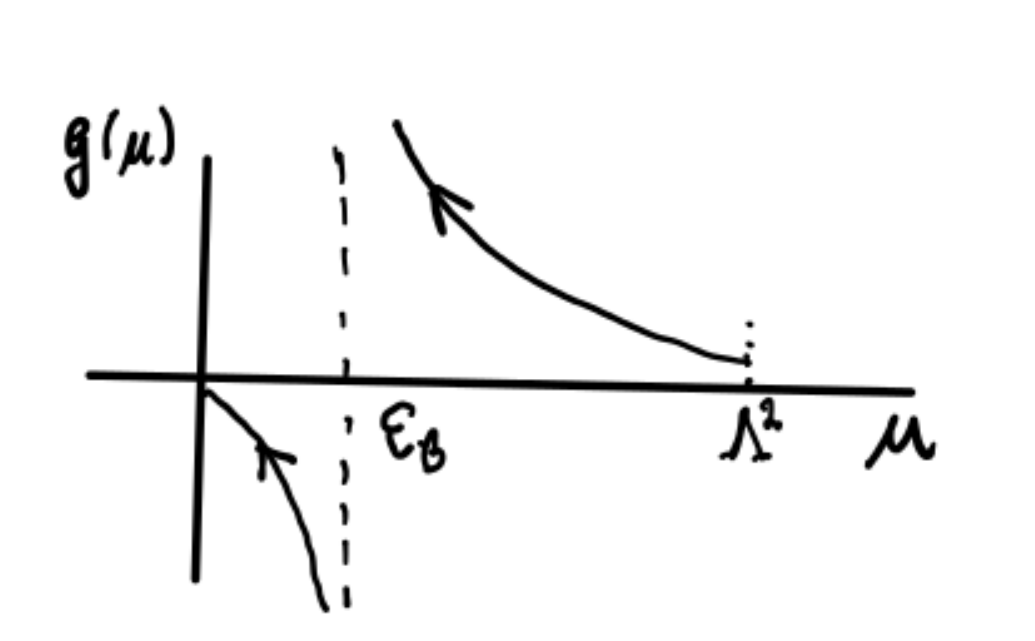}\\
\caption {  The flow of  $g $ as  the
 energy scale $\epsilon$  decreases. The infrared Landau pole occurs at energy $\epsilon_B$. 
 }   \label{fg-2}  
 \end{figure}

The renormalization group flow is depicted in figure \ref{fg-2}. 
The running coupling (\ref{beta1}) has  a Landau pole in the infrared.  The pole occurs at the energy scale
\begin{align}
\label{landau_pole}
\epsilon_B = \Lambda^2 \exp\left( -\frac{8\pi N}{g^2(\Lambda)}\right)  
\end{align}
which is exponentially small in $N$. This pole in the two-body scattering amplitude is due to a bound state in the Fermion-Fermion channel, the Cooper pair,
which must always be present for an attractive two-body potential in two dimensions.   The fact that there is Cooper pair formation in our system which will
have non-zero density means that the final fate of the ideal system should be a superfluid condensate of Cooper pairs.   However the pair is exceedingly
weakly bound and the BCS gap which would form is so small that any randomizing effect, or a small temperature would destroy this condensate.  There will
be plenty of space between the scale of the Cooper pair and the scale of the density, which will occur at a much larger scale, by a factor with an exponential
in $N$, where such Cooper pair destroying effects could occur.  Moreover, if Cooper pair condensation did occur, it could only happen when the density is nonzero and it would then only serve to lower the energy of the state that we shall find.

 To find the grand canonical potential to the next-to-leading order in 1/N, 
it is convenient to begin with the action 
\begin{align}\label{tildeaction}
\tilde S = \int dtd^2x\tilde{\mathcal L} 
\end{align}
obtained from integrationg the Lagrangian density
\begin{align}
\tilde{\mathcal L} = \psi_a^*\dot\psi_a+\vec\nabla\psi_a^*\cdot\vec\nabla\psi_a
 -(\mu+ g\sigma)\psi_a^*\psi_a
  +\frac{{\rm  N}g\sigma^2}{2}  
\end{align}
and  the functional integral
\begin{equation}\label{pathintegral}
 e^{-\Phi(\mu,g,N){\mathcal V}{\mathcal T}}=\int [d\psi_a(x) d\psi^*_a(x)d\sigma(x)] ~e^{-\tilde S[\psi,\psi^*,\sigma] }
\end{equation}
The Hubbard-Stratonovich $\sigma$-field has need introduced in order to facilitate the large $N$ expansion.  
It  is  related to the Fermion density per species, for example, its expectation value
is the density per species defined in equation (\ref{rho})
$$
\left< \sigma(x)\right>=   \rho $$ 
and its connected correlation function differs from the density-density correlation function  by a trivial constant
$$
<\sigma(x)\sigma(y)>_C=<\rho(x)\rho(y)>_C-\frac{1}{g}\delta(x-y)
$$

If we begin with the functional integral in equation (\ref{pathintegral}) and if we 
first do the Gaussian functional integral
over  $\sigma(x)$,  we recover the purely Fermionic theory of equation (\ref{lagrangian_density}).  
If, instead of integrating $\sigma(x)$, we first perform 
the Gaussian integral over $\psi_a(x)$, we obtain the effective action for the $\sigma(x)$-field
\begin{equation}
S_{\rm eff}=  \int dtd^2x \frac{{\rm  N}g\sigma^2}{2}  
-{\rm  N}{\rm Tr}\ln(\partial_t-\nabla^2-(\mu+g\sigma))
\end{equation}
This effective action is proportional to N and the dynamics of $\sigma(x)$ are therefore weakly coupled 
and semi-classical when N is large.  
The remaining functional integration over the variable $\sigma(x)$ can therefore  be computed  in the 
 saddle point approximation.  
To proceed, we
 denote $$\sigma(x)=\rho+\delta\sigma(x)$$
 where, we recall that the expectation value $<\sigma(x)>=\rho$.     
  Then, to get the next to leading order, the order $N^0$ terms in the grand canonical potential, 
  according to the usual prescription for the background field technique  \cite{Jackiw:1974}, 
  we expand $S_{\rm eff}$ to quadratic order in $\delta\sigma(x)$,  we drop the linear terms,  
 and we do the Gaussian integral over $\delta\sigma(x)$.  The result is
 the potential density
 \begin{align}
\Phi(\mu,g,N)=&{\rm Inf}_\rho  \frac{{\rm  N}g\rho^2}{2}  
-\frac{\rm  N}{\mathcal V \mathcal T}{\rm Tr}\ln D^{-1} 
+\frac{1}{2\mathcal V\mathcal T}{\rm Tr}\ln \Delta^{-1}
+\ldots
\label{effective_action}\end{align}
where  the ellipses denote terms of order $\frac{1}{N}$ and 
\begin{align}
&\Delta^{-1}(x,y)=N\left[g\delta(x,y)+g^2\Pi(x,y)\right]\label{PI}
\\
&
\Pi(x,y)=D(x,y)D(y,x)
\label{S}\\
&
D(x,y)=(x|\frac{1}{\partial_t-\nabla^2-(\mu+g\rho)} |y) \label{D}
 \end{align}
 The variable $\rho$ is to be determined so that it minimizes the function on the right-hand-side of 
 equation (\ref{effective_action}).   The value of $\rho$ at the minimum of the physical density per species defined in equation (\ref{rho}). 
 The value of the right-hand-side at its minimum is the thermodynamic potential $\Phi(\mu,g,N)$.

 If we  keep only the leading
 order in N (the first two terms in (\ref{effective_action})), do the simple Feynman integral which computes the second term 
 and cancel terms which diverge like a positive powers of the cutoff
 with counter-terms (there are no logarithmic divergences 
 at this order), this approximation to the potential   (\ref{effective_action}) is equal to  equation  (\ref{infinite_N}) whose
 implications we have already discussed in the text that followed  equation  (\ref{infinite_N}).   
 
  To proceed to the next order of the 1/N expansion, consider the integral which evaluates the last term on the right-hand-side of
equation  (\ref{effective_action}).  
 \begin{align}
  \frac{1}{2\mathcal V\mathcal T}{\rm Tr}\ln\Delta^{-1}=\frac{1}{2}  \int \frac{d\omega}{2\pi}\int_0^{\Lambda^2}\frac{dk^2}{4\pi}\ln \Delta^{-1}( \omega,  k^2)  
 = \left[ -\frac{g^2 (\mu+g\rho)^2}{256\pi^3}\ln
\frac{\Lambda^2}{\mu+g \rho}
+(\mu+ g\rho)^2\frac{\varphi(g)}{4\pi}+\ldots\right]
\end{align}
where we have isolated the quartic, quadratic and logarithmically divergent terms, 
dropped the quartic and quadratically divergent ones, assuming that they are 
canceled by counter-terms that we would add to the original Lagrangian density and we name the
remaining finite integral  $( \mu+g\rho)^2\varphi(g)/4\pi$. Since it is finite, the  $ \mu+g\rho$-dependence
of the term in which it occurs is
determined by dimensional analysis.  Our results will not depend on the precise form of the function $\varphi(g)$. 

The renormalization of this theory  is well-known and in the leading order it is confined to
coupling constant renormalization which leads to the beta function (\ref{beta}).    What is more, even after renormalization, 
the remaining logarithmic terms spoil the
large N expansion.  When evaluated on the solutions for $\rho\sim e^{-N...}$,  the logarithms are large and they compensate the small factors of $1/N$.  
This is a well-known phenomenon
which already occurs in the fist example in the paper by Coleman and Weinberg \cite{Coleman} on dynamical symmetry breaking.  
The troublesome terms are of the form $\sim\left( \frac{1}{N}\ln \frac{\rho}{\mu}\right)^k$ and they are not suppressed at large $N$ when 
$\ln\rho\sim N$. The solution of this problem
is to use the renormalization group to re-sum these logarithmically singular terms to all orders.  The result is simple.  It is given by replacing  
  $g$ in the potential by $g(\mu+g\rho)$, the running coupling evaluated at the scale of the effective Fermi energy, $\mu+g\rho$. 
Then, to the renormalization group improved potential at the next-to-leading order in the large N expansion is
 \begin{align}
\Phi(\mu,g,N) =N~{\rm Inf}_\rho ~ \left[ \frac{g\rho^2}{2}  -\frac{(\mu+g\rho)^2  } {8\pi}\left(1-\frac{\varphi(g)}{\rm N}\right)  
+\right.  \nonumber \\
.\left. +g\left(\rho-\frac{\mu +g\rho}{4\pi}\right)^2\frac{g^2}{16\pi {\rm N}}\ln\frac{\Lambda^2}{\mu+g\rho}
 +{\mathcal O}\left(\frac{1}{N^2}\right) \right]
\label{final} \end{align} 
 The cutoff dependence has not entirely canceled.  However, the  cutoff-dependent term is
equal to the square of the derivative by $\rho$ of the leading order terms and it and its derivative will vanish to the order
in 1/N to which we are working  when it is evaluated on the 
solution of the equaton  for $\rho$.  
The equation  $\frac{\partial\Phi}{\partial\rho}=0$ yields  
 \begin{align}
\mu =4\pi \rho\left[  
  1+\frac{\varphi(g(4\pi\rho)) }{\rm N}-\frac{ \beta(g(4\pi\rho))}{2g(4\pi\rho)}-\frac{g(4\pi\rho)}{4\pi} \right] + {\mathcal O}\left(\frac{1}{N^2}\right)
\label{rho111} 
\end{align}
This equation determines the relationship between $\rho$ and $\mu$.   Its solution for $\rho$ gives the physical density corresponding
to a given chemical potential $\mu$.  
Remembering that $\beta\sim1/$N, we can see that (\ref{rho111}) is identical to (\ref{lng-1}) with additional order 1/N corrections.
Plugging (\ref{rho111})  into $\Phi$ in (\ref{final}), we obtain the physical value of the grand canonical potential. Remembering that, on the solution, 
the pressure is given by $P=-\Phi$, yields the pressure,
\begin{align}
P=2\pi\rho^2 {\rm N} \left[  
   1+\frac{\varphi(g(4\pi\rho) ) }{\rm N}- \frac{\beta(g(4\pi\rho)}{g(4\pi\rho) } -\frac{g(4\pi\rho) }{4\pi} \right] 
 +{\mathcal O}\left(\frac{1}{N}\right)
\label{final1} \end{align}

\begin{figure}[h]
\centering
\includegraphics[width=7cm,clip]{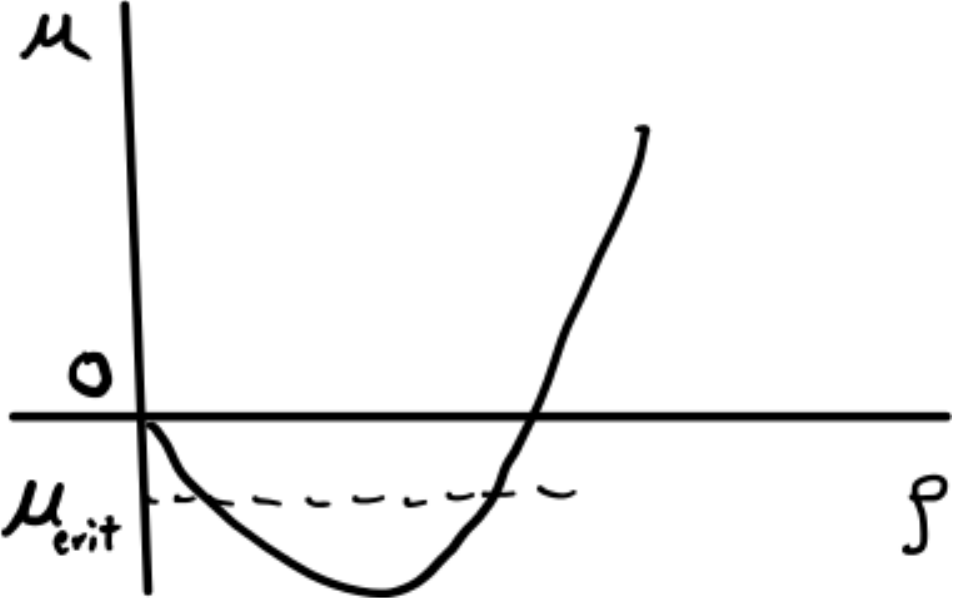}\\
\caption {  A schematic diagram of the chemical potential, $\mu$, plotted on the vertical axis and the 
density $\rho$ plotted on the horizontal axis.   The horizontal dotted line represents the chemical potential where
the pressure vanishes. There are two values of $\rho$ for which this occurs and only the larger one is stable.  The criterion for stability
is that the slope should be positive at the solution. }   \label{fg-4}  
 \end{figure}  
 
When $\mu$ is negative, there are two solutions of equation  (\ref{rho111}) for $\rho$ as a function of $\mu$ and we must decide
which solution is the stable one by comparing their grand canonical potential densities.   A schematic of the locus of (\ref{rho111}) in 
the $\mu$-$\rho$ plane is shown in figure \ref{fg-4}.    
The fact that there are two solutions is easy  to see analytically when $\mu=0$.  In that case, one solution is
$\rho=0$, which has zero pressure, $P=0$.  
The other solution is $\rho$ such that $g(4\pi\rho)$ solves the equation  
$$
1+\frac{\varphi(g(4\pi\rho)) }{\rm N}-\frac{ \beta(g(4\pi\rho))}{2g}-\frac{g(4\pi\rho)}{4\pi}=0
$$  which, as a function of $g$, has a solution $g(4\pi\rho)= 4\pi+{\mathcal O}(1/{\rm N})$. 
The pressure, evaluated at this solution, 
$$
P(\mu=0)= 2\pi\rho^2 {\rm N} \left[  - \frac{\beta(g(4\pi\rho)}{2g(4\pi\rho) }  \right] 
 +{\mathcal O}(1/N)
 $$  
 is  positive (because $\beta$ is negative).  This means that this second  solution, with non-zero density, is the 
 preferred state.\footnote{Remember that we are minimizing $\Phi$. The local minimum with smallest $\Phi$ is
 the one with largest pressure.} Thus, at $\mu=0$ the state with finite density is the stable one.  However, now that
 scale symmetry is intrinsically broken by the non-zero beta function, there is nothing special about the point $\mu=0$.
 This is not the critical point. 
 
  We could imagine further decreasing the density by 
 lowering $\mu$ to smaller values -- smaller than zero means to negative values  -- until 
 we reach the critical value of the chemical potential, $\mu_{\rm crit}$.  At this point,  the finite density state becomes degenerate with
 the zero density state.  All of the correction terms in the grand canonical potential (\ref{final}) assumed that the density is nonzero.  If the density is zero,
 the potential and the pressure are both zero.
 This occurs when we reach at a lower
limit of the density which we shall call $\rho_{\rm crit.}$.  

A schematic plot of $\mu$ versus $\rho$ is shown in figure \ref{fg-4}. There, we see that
when $\mu$ becomes negative, just below zero, there are two values of $\rho$ which correspond to the same value of $\mu$.  We must chose
the larger one as it is the one with the positive slope.  The sign of the slope determines the sign of the compressibility (or the curvature of the grand 
canonical potential) and the solution is stable if this slope is positive.   We can continue to lower $\mu$ until we either come to a state where the
pressure vanishes, and the state is degenerate with the state with zero density, or we come to the bottom of the
curve, where the compressibility diverges. In this case, we come to the state where the pressure vanishes first.  (We can show this analytically.)

At that point, the chemical potential  $\mu_{\rm crit.}$ is small and negative.  
The density is small and still positive, $\rho_{\rm crit.}$. The vanishing of the pressure, with with non-zero density occurs
when the coefficient in equation (\ref{final1}) for the pressure 
$$
\left[  
   1+\frac{\varphi(g(4\pi\rho) ) }{\rm N}- \frac{\beta(g(4\pi\rho)}{g(4\pi\rho) } -\frac{g(4\pi\rho) }{4\pi} \right] 
   $$
vanishes.  
We can think of this as an equation for a critical coupling constant $g^*$, 
 which is 
\begin{align}\label{gcrit}
 1+\frac{\varphi(g^*) }{\rm N}- \frac{\beta(g^*)}{g^*}-\frac{g^*}{4\pi} +{\mathcal O}(1/{\rm N}^2)=0
 \end{align}
 This condition is solved by the critical value of the coupling
 \begin{align}
 g^*=4\pi\left( 1+\frac{\varphi(4\pi) }{\rm N}- \frac{\beta(4\pi)}{4\pi}+{\mathcal O}(1/{\rm N}^2)\right)
 \label{gstar}
   \end{align}
   which is the same critical coupling that we found in the infinite $N$ limit, corrected by terms of order $\frac{1}{N}$. 
   This is a critical coupling in the sense that the running coupling constant evaluated at the scale  
    $4\pi\rho_{\rm crit.}$ must be equal to this critical coupling:
      \begin{align}\label{rhocrit1}
g(4\pi\rho_{\rm crit.})=g^*
\end{align}

The critical chemical potential indeed turns out to be of order 1/N and negative, 
 \begin{align}
 \mu_{\rm crit.}= 2\pi\rho_{\rm crit.}\frac{\beta(g^*)}{g^*}=-\frac{\pi}{N}\rho_{\rm crit.}
 \end{align}  
 The critical value of the density is
 \begin{align}\label{rhocrit}
 \rho_{\rm crit.} = \Lambda^2 e^{-8\pi N\left( \frac{1}{g(\Lambda^2)}-\frac{1}{g^{*}}\right)}
\end{align}

We can see from (\ref{rhocrit}) that,   to create a hierarchy of scales, so that $\rho_{\rm crit.}<<\Lambda^2$ we only need 
the renormalization group trajectories such that the coupling at the ultraviolet scale lies in the band 
$0\leq g(\Lambda^2)<g^* $.  Then, the renormalization group flow will self-tune the coupling for us.

Once the theory is no longer exactly scale invariant, the 
compressibility no longer diverges at the critical point, but it is large, of order N times what it would be for an ideal gas, 
\begin{align}
\kappa  = \kappa_0 \frac{2N+\ln\frac{\rho}{\rho_{\rm crit.}} }{ 1+\ln \frac{\rho}{\rho_{\rm crit.}} }
\end{align}

The damping constant of the dilaton at the critical density is obtained from the curvature of the potential, 
$ \frac{\partial^2\Phi}{\partial\rho^2}$ at $\rho=\rho_{\rm crit.}$.  Again, once the theory is no longer scale invariant,
the damping  no longer goes to zero.  However, at the critical density, it is still small, of order $\frac{1}{N}$ and if $N$ is 
large enough, the dilaton is still a long-lived excitation.   .  
Equation  (\ref{ss}) is replaced by
\begin{align}\label{ss1}
\Delta(\omega,k)=\frac{\frac{1}{{\rm N}\pi}  \sqrt{4\pi \rho_{\rm crit.} }|\vec k|}{\Gamma-i\omega}
~,~
\Gamma=  \frac{1 }{{\rm N}}    \sqrt{4\pi \rho_{\rm crit.}   }|\vec k| 
\end{align}
The pressure and compressibility are positive, indicating stability of this phase
for values of $\mu$  larger than $\mu_{\rm crit.}$.

 \begin{figure}[h]
\centering
\includegraphics[width=7cm,clip]{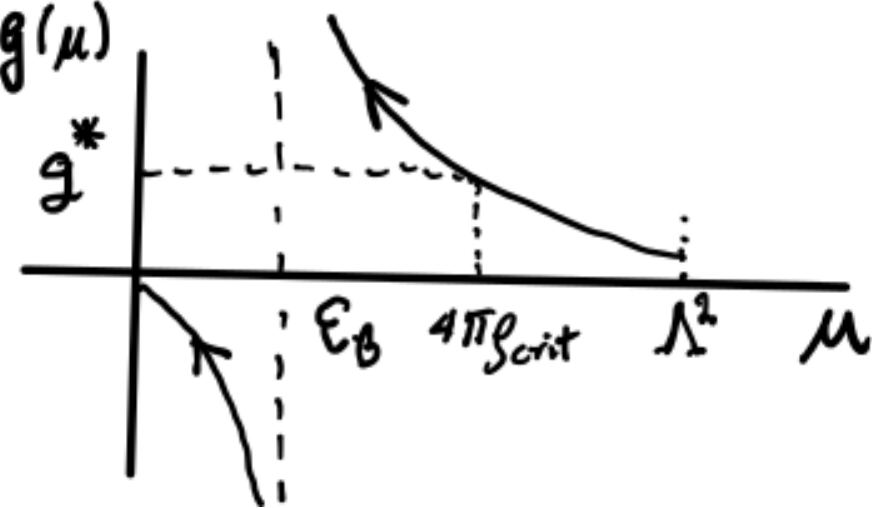}\\
\caption {  The flow of  $g $ as  the
 energy scale $\epsilon$  decreases. 
The critical minimum density is determined by Eqs.\  (\ref{rhocrit1}) and (\ref{gstar})  which
always has a solution if $0<g(\Lambda^2)<g^*$. 
The dashed line is the position of the Landau
pole which is exponentially smaller than $\rho$ in
the large N limit (see equation (\ref{landau_pole})).
 }   \label{fg-3}  
 \end{figure}

\begin{figure}[h]
\centering
\includegraphics[width=7cm,clip]{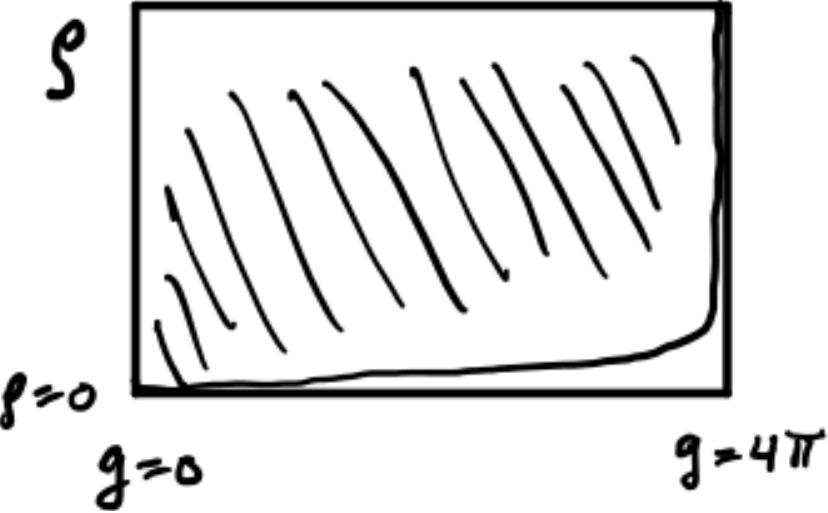}\\
\caption {   The phase diagram.  The density is nonzero above and to the left of the critical line.  As we either lower the density, or increase the coupling, we 
approach the critical line where the density suddenly jumps to zero.   In the infinite $N$ limit, the critical curve is packed onto the corner, so that when the coupling is less than $g^*=4\pi$, the critical density is at zero snd the phase transition is second order.  Then, when $g=g^*$, the density can take any arbitrary value, and the quantum phase transition at $g=g^*$ is also second order.   Then $N$ is lowered from infinity, the critical curve moves away from the corner and becomes the line that is shown.  The phase transition becomes a fluctuation induced first-order transition.  The phase below the line either has zero density, in the grand canonical ensemble, or it consists of a mixed phase of zero density and clumps of atoms.).
 }   \label{fg-5}  
 \end{figure}

  The first order behaviour arises from the fact that, as  seen in figure \ref{fg-3}, 
  the density is always larger than the scale of 
the  Landau pole which occurs at 
\begin{align}\label{landau_pole}
\epsilon_B= \Lambda^2 e^{- \frac{8\pi N}{g(\Lambda^2)}  } \sim  e^{-  2N} \rho_{\rm crit.}
\end{align}
The phase diagram is depicted in figure \ref{fg-5}

\section{Epilogue}

The position of the infrared Landau pole coincides with the binding energy of a Cooper pair which 
appears in the Fermion-Fermion channel.
This weakly bound pair should Bose condense at zero temperature so that, in the strict mathematical sense,
the final state of this system should be a superfluid.  This Bose condensation is only possible if there is a density of
Fermions to pair and it should therefore aid the stability of the symmetry breaking phase by further lowering its free energy. 
In that case, the superfluid condensate $<\psi_a\psi_b>$ transforms in the $\left(\begin{matrix} N\cr 2\cr \end{matrix}\right)$-dimensional
second fundamental representation of SU(N).  It should break the global symmetry from $U(N)$ to $U(2)\otimes U(N-2)$.  

However, the Cooper pair  binding energy is exceedingly small, exponentially smaller than the density, from equations (\ref{landau_pole})
and (\ref{rhocrit}), we see that
\begin{align}\label{hierarchy}
\rho_{\rm crit.}=e^{2N}\epsilon_B>>\epsilon_B
\end{align}
Since the cooper pair binding is so weak, 
even a   small temperature  in the range
$\epsilon_B<k_BT<\rho$ or other small randomizing effects 
would destabilize the superfluidity but not necessarily the spontaneous density. In either case, it would be interesting to understand
further properties of the large $N$ gas.

 Although the system that we have described is largely paradigmatic, it could conceivably 
 be experimentally realizable, for example,  
using Fermionic atoms with high spin and an approximately spin independent attractive short-ranged interaction so that, at least
its effective field theory is described by (\ref{lagrangian_density}).  Such systems are already realized for N up to N=10 \cite{needs} \cite{desalvo}, which should
be large enough to create a hierarchy of scales $\rho_{\rm crit.}=e^{2N}\epsilon_B>>\epsilon_B$. 
The interaction would have to be attractive  and the coupling at the ultraviolet scale of this effective field theory would have lie in the appropriate intertval..

  What we have done in the above should be accurate if $N$ is large enough, but how large is large  enough is not entirely clear and not easy to estimate from only the leading and next-to-leading order of the computation.  A related question is,  if $N$ were not really large, would the phenomenon that we describe  persist.   This question could be answered by numerical simulations where the technology for looking at $N=2$ already exists   \cite{anderson} and it should be extendable to large $N$. 
    It could tell us whether, for example. somewhere at very strong coupling, there is a phase with approximate scale symmetry even when $N$ is small.    
  
  \noindent
  This work is supported in part by NSERC. The author thanks Fei Zhou and Arkady Vainshein for discussions.

\end{document}